\documentstyle[twocolumn,,aps,psfig]{revtex}

\begin{document} 

\title{Influence of the Environment Fluctuations on Incoherent Neutron 
Scattering Functions}

\author{D. J. Bicout}

\address{Institut Laue-Langevin, Theory Group\\
6, Rue Jules Horowitz, B.P. 156 \\  
38042 Grenoble Cedex 9, France}

\date{\today}

\maketitle

\begin{abstract} 
In extending the conventional dynamic models, we consider a simple model to 
account for the environment fluctuations of particle atoms in a protein 
system and derive the elastic incoherent structure factor (EISF) and the 
incoherent scattering correlation function $C(Q,t)$ for both the jump 
dynamics between sites with fluctuating site interspacing and for the 
diffusion inside a fluctuating sphere. We find that the EISF 
of the system (or the normalized elastic intensity) 
is equal to that in the absence of fluctuations averaged over the 
distribution of site interspacing or sphere radius $a$. The scattering 
correlation function is 
$C(Q,t)=\sum_{n}\langle{\rm e}^{-\lambda_n(a) t}\rangle\,\psi(t)$, 
where the average is taken over the $Q$-dependent effective distribution of 
relaxation rates $\lambda_n(a)$ and $\psi(t)$ is the correlation function 
of the length $a$. When $\psi(t)=1$, the relaxation of $C(Q,t)$ is exponential 
for the jump dynamics between sites (since $\lambda_n(a)$ is independent 
of $a$) while it is nonexponential for diffusion inside a sphere. 
\end{abstract}

\begin{center}
{\bf PACS numbers:} 02.50.Ey ; 66.10.Cb ; 87.15.He 
\end{center}

\newcounter{xeq}

By analyzing the incoherent scattering function, $S({\bf Q},\omega)$,
where ${\bf Q}$ is the scattering wavevector and $\hbar \omega$ the energy 
transfer, techniques of quasielastic neutron scattering from hydrogen atoms  
(the main neutron scatterers in a typical protein) allow to study motions of 
particles (atoms, molecules, chemical species, ...) in biological systems 
\cite{Bee}. For this purpose, there are many physical situations of interests 
like molecules at surfaces, in micellar systems or in vesicles and structural 
cages, in which the system is modeled by jump of particles among sites or by 
diffusion inside a confining geometry \cite{Vol,Hall,Bel,BelB}. For instance, 
when the problem can be described by the jump dynamics of a particle between 
two non-equivalent sites separated by a distance $a$, the elastic part of 
$S(Q,\omega)$ (i.e., the elastic incoherent structure factor (EISF)), is 
given by 
\begin{equation}
A_0(Qa)=1-2p\,(1-p)\,\left[1-j_0(Qa)\right]\:,
\label{A0j}
\end{equation}
where $j_{l}(z)$ is the spherical Bessel function of the first kind of 
order $l$ and $p$ is the probability of finding the particle in one of the 
sites. The incoherent scattering correlation function $C(Q,t)$ (i.e., 
the inverse Fourier transform of the quasielastic part of $S(Q,\omega)$) is 
a single exponential independent of $Q$ given by $C(t)={\rm e}^{-\Gamma t}$ 
where $\Gamma$ (related to mean residence times of the particle in either 
site) is the relaxation rate of the probability of finding the particle in 
either site. This jump model has been used to study the internal molecular 
motions and to interpret the dynamical transition in 
proteins \cite{Doster,Fitter}. 

Likewise, when the problem can be described by the isotropic 
diffusion inside a sphere with impermeable surface \cite{Vol}, one finds that 
for $Q$ small, the quasielastic part of $S(Q,\omega)$ is well described by a 
single Lorentzian with the linewidth $\Gamma_0=4.33 D/a^2$, where $a$ is the 
sphere radius and $D$ the diffusion constant of the particle. Equivalently, 
$C(Q,t)\simeq {\rm e}^{-\Gamma_0 t}$ where $\Gamma_0^{-1}$ is the typical 
time for a particle to diffuse over the entire sphere. For times of order 
or larger than $\Gamma_0^{-1}$ (i.e., in the $\omega\ll \Gamma_0$ limit) the 
EISF is given by 
\begin{equation}
A_0(Qa)=|F(Qa)|^2=\left|\frac{3j_1(Qa)}{Qa}\right|^2\:.
\label{FQ}
\end{equation}
Such an analysis is used by several authors \cite{Bel,Per,Den} to study, for 
instance, the internal dynamics, structure and dynamics of surface molecules 
in proteins. In these studies, the particles (mainly hydrogen atoms) are 
assumed to diffuse within 
permanent spherical cages.

On the other hand, it is well known that proteins are fluctuating systems 
which undergo configurational fluctuations in their structure. In different 
conformational substates, a protein may have the same coarse structure 
but differs in local configurations leading to the fluctuations of
the local environment of each protein atom. As a result, a distance between 
sites or a structural cage in a protein fluctuate in length or size (and shape) 
and have a finite lifetime due to local structural relaxation just like in 
mode-coupling picture of liquids \cite{Gotze}. In this respect, neglecting 
the fluctuations of shapes for simplicity, the dynamics of each hydrogen 
atom in a typical protein can be described, in the first approximation, as 
a jump dynamics between sites separated by fluctuating distances or by 
diffusion inside a fluctuating sphere. 

In this paper, we present a simple model analysis to account for fluctuations 
of the local environment. We focus on the derivation of the EISF and $C(Q,t)$ 
in situations in which the site interspacing $a$ (for the jump between sites) 
or the sphere radius $a$ (for the diffusion inside a sphere) is allowed to 
fluctuate in the course of time. To model simply the fluctuations of $a$, we 
consider the following dynamics for the system formed of a particle atom 
jumping between two sites (or diffusing inside a sphere with impermeable 
surface): the particle keeps jumping between sites separated by $a(t)$ 
(or diffusing within the spherical cage of radius $a(t)$) until it suffer a 
{\em configurational collision} of zero duration that equilibrates both 
the site interspacing (or the sphere radius) and the particle position. That 
is to say that after a collision the update particle position and 
the site interspacing (or the sphere radius) are chosen according to the 
normalized equilibrium distribution, 
\begin{equation}
P_{\rm eq}({\bf r},a)=
\frac{{\rm e}^{-\beta [V(a)+U({\bf r},a)]}
}{\int_{0}^{\infty}\!da\,{\rm e}^{-\beta V(a)}\,
\int_{0}^{4\pi}\!d\Omega \int_{0}^{\infty}\!r^2\,
{\rm e}^{-\beta U({\bf r},a)}\,dr}
\label{peq}
\end{equation}
where ${\bf r}=(r,\Omega)$, $r$ is the particle position, $\Omega$ stands 
for polar and azimuthal coordinates, $U({\bf r},a)$ is the potential energy 
for the particle position for a fixed $a$ while $V(a)$ is the potential of 
free energy associated to the length (spherical cage of radius) $a$, and 
$\beta^{-1}={\rm k_BT}$ is the thermal energy. The reduced equilibrium 
distribution for the length $a$ is defined as, 
$p_{\rm eq}(a)=\int_{0}^{4\pi}\!d\Omega \int_{0}^{\infty}\!r^2\,
P_{\rm eq}({\bf r},a)\,dr$.
The waiting time between successive configurational collisions is a random 
variable with distribution $\phi(t)$ related to the stationary correlation 
function of site interspacing (or the sphere radius) by 
\begin{equation}
\psi(t)=\frac{\langle a(t)a(0)\rangle}{\langle a^2\rangle}=
\int_{t}^{\infty}\phi(\tau)\,d\tau\:.
\label{radcor}
\end{equation}
Assuming that the environment fluctuations act on the particle dynamics in 
renormalizing just the residence times (for jump dynamics) or the 
diffusion constant (for diffusion dynamics), thus the Green's function, i.e. 
the probability density of finding the particle at the position ${\bf r}$ with 
site interspacing $a$ (inside a sphere of radius $a$) at time $t$ given that 
it was initially at ${\bf r}_0$ with site interspacing $a_0$ (inside a 
sphere of radius $a_0$), is given by   
\begin{eqnarray}
G({\bf r},a,t|{\bf r}_0,a_0) & = &G_0({\bf r},t|{\bf r}_0;a)\,\delta(a-a_0)\,
\psi(t)\nonumber \\
& &+\,P_{\rm eq}({\bf r},a)\,\left[1-\psi(t)\right]\:,
\label{green}
\end{eqnarray}
where $G_0({\bf r},t|{\bf r}_0;a)$ is the Green's function for a particle 
with a fixed site interspacing $a$ (or diffusing inside a sphere of a fixed 
radius $a$). 

The incoherent intermediate scattering function, \\ $I({\bf Q},t)$, (which 
is the inverse Fourier transform of $S({\bf Q},\omega)$) is 
$I({\bf Q},t)=$ \\ $
\int da_0\int d{\bf r}_0\int da\int d{\bf r}\,{\rm e}^{i{\bf Q.r}}\,
G({\bf r},a,t|{\bf r}_0,a_0)\,
{\rm e}^{-i{\bf Q.r}_0}\,P_{\rm eq}({\bf r}_0,a_0)$. Using into this relation 
the expression of the Green's function in Eq.(\ref{green}), one can show that 
for an isotropic problem the function $I({\bf Q},t)$ splits into two parts as, 
\begin{equation}
I(Q,t)=A(Q)+\left[1-A(Q)\right]\,C(Q,t)\:,
\label{IQ0}
\end{equation} 
where the elastic part, i.e. the EISF of the system, is
\begin{equation}
A(Q)=\left|\int_{0}^{\infty}\!da\int_{0}^{4\pi}\!d\Omega 
\int_{0}^{\infty}\!r^2\,{\rm e}^{i{\bf Q.r}}\,P_{\rm eq}({\bf r},a)\,dr
\right|^2\:,
\label{AofQ}
\end{equation} 
and the incoherent scattering correlation function $C(Q,t)$ containing all 
information about the relaxation dynamics, is 
\begin{eqnarray}
C(Q,t) & = &\left\langle \sum_{n=1}^{\infty}
g_n(Qa)\,{\rm e}^{-\lambda_n(a)\,t}\right\rangle\,\psi(t)\nonumber \\
& = & \left[\sum_{n=1}^{\infty}\int_{0}^{\infty}\!da\,
{\overline g}_n[\beta V(a),Qa]\,{\rm e}^{-\lambda_n(a)\,t}\right]\,\psi(t)\:,
\label{CQt}
\end{eqnarray}
where the average of any function $f(a)$ is defined as 
$\langle f(a)\rangle=\int_{0}^{\infty}\!p_{\rm eq}(a)\,f(a)\,da$. The 
$g_n(Qa)$ is the $Q$-dependent bare distribution of relaxation 
rates $\lambda_n(a)$ in the absence of fluctuations while 
${\overline g}_n[\beta V(a),Qa]$ is the $Q$-dependent effective distribution 
of relaxation rates which accounts for distribution of $a$.
To be specific, we assume that the effective potential of free energy 
associated to the length $a$ is given by
\begin{eqnarray}
V(a)=\left\{\begin{array}{lcl}
\sigma a^2-\varepsilon-2{\rm k_BT}\,\ln(a) & ; & a\geq R\:,\\
\infty & ; & a<R\:.
\end{array}\right.
\label{pot}
\end{eqnarray}
where $\varepsilon=\sigma R^2$, $R$ (of order of the Van der Waals contact 
length) is the minimum value of $a$, $\sigma$ is the force constant and the 
repulsive logarithmic term represents the entropic contribution which accounts 
for the increase in configuration space as $a$ increases. As a result of 
the balance between the attractive harmonic and repulsive entropic terms, 
the length $a$ has an equilibrium value at finite temperature $T$ given by 
$a_{\rm eq}=\mbox{max}\left[R,({\rm k_BT}/\sigma)^{1/2}\right]$. We note in 
passing that such a  harmonic potential in Eq.(\ref{pot}), without the 
entropic term, has been used to study the hydrophobic effect in the volume 
fluctuations of globular proteins \cite{Lee}. 

In what follows we derive in detail the EISF and the $C(Q,t)$ in 
Eqs.(\ref{AofQ}) and (\ref{CQt}), respectively, for the jump dynamics 
between sites and diffusion inside a sphere.

\section*{Fluctuating Site Interspacing}

\begin{figure}[ht]
\vspace{0.3cm}
\centerline{\psfig{figure=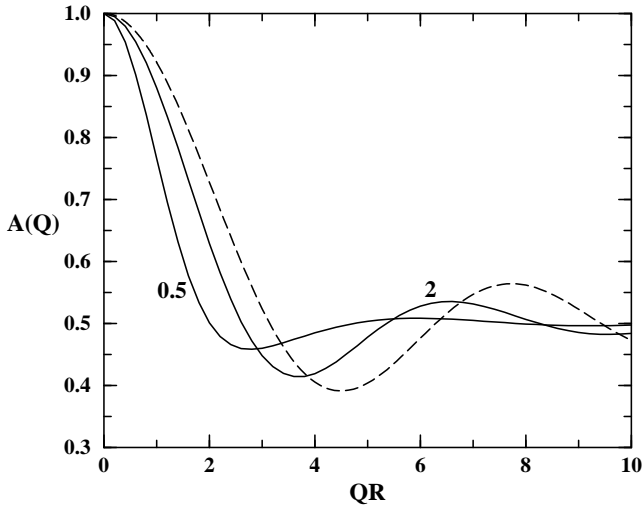,width=3.3in,angle=0}}
\vspace{0.25cm}
\caption{The EISF $A(Q)$ in Eq.(\ref{ACjump}) as a function of 
$QR$ for $p=0.5$ and $\varepsilon/{\rm k_BT}=0.5,2$ as quoted on the figure. 
The dashed line represents $A_0(Q)$ in Eq.(\ref{A0j}) for $a=R$ and $p=0.5$ 
for comparison.} 
\label{Ajump}
\end{figure}

Consider the jump dynamics of a particle between two non-equivalent sites 
separated by a fluctuating distance $a$ of stationary correlation function 
$\psi(t)$ and with $\tau_1$ and $\tau_2$ being the effective mean residence 
time of the particle in each site. For one site located at the origin, we have 
\begin{eqnarray}
{\rm e}^{-\beta U({\bf r},a)}=\left\{\begin{array}{lcl}
p & ; & {\bf r}={\bf 0}\:,\\
1-p & ; & {\bf r}={\bf a}\:,\\
0 & ; & \mbox{otherwise}\:,
\end{array}\right.
\:;\:p=\frac{\tau_1}{\tau_1+\tau_2}\:.
\label{potj}
\end{eqnarray}
The distribution of relaxation rate is a delta function, 
$g_n(Qa)={\overline g}_n[\beta\varepsilon,Qa]=\delta_{n,1}$ with 
$\lambda_1=\Gamma=\tau_1^{-1}+\tau_2^{-1}$. In this case, the normalized 
scattering intensity is given by Eq.(\ref{IQ0}) with the EISF and the 
scattering correlation function (independent of $Q$) given by 
\begin{equation}
A(Q)=\langle A_0(Qa)\rangle\:\:\:\mbox{and}\:\:\:
C(t)={\rm e}^{-\Gamma t}\,\psi(t)\:,
\label{ACjump}
\end{equation}
where $A_0(Qa)$ is given in Eq.(\ref{A0j}). Since $\Gamma$ is independent of 
the site interspacing, thus the effect of fluctuations on $C(t)$ is simply 
multiplicative as in (\ref{ACjump}). 
As a result of site interspacing fluctuations, the structure factor $A(Q)$ 
(shown in Fig.~\ref{Ajump}) is the EISF for the jump dynamics between two 
sites separated by a fixed distance $a$ averaged over the distribution of 
the distance. As $\beta\varepsilon$ gets smaller, the extrema of $A(Q)$ 
become less pronounced and are shifted to the left compare to $A_0(Q)$. 
The mean-squared displacement of the particle is 
\begin{eqnarray}
& & \langle X^2\rangle=\left.-3\,
\frac{d\ln\left[A(Q)\right]}{d(Q^2)}\right|_{Q=0}=
p\,(1-p)\,\langle a^2\rangle \nonumber \\
& = & p\,(1-p)\,\left[\frac{3R^2}{2\beta\varepsilon}
+\frac{R^2}{1+(\pi/2\beta\varepsilon)^{1/2}\:{\rm e}^{\beta\varepsilon}\,
{\rm erfc}\left(\sqrt{\beta\varepsilon}\right)}\right]\:.
\label{x2jump}
\end{eqnarray}
Solid line in Fig.~\ref{roft}B represents this function. When 
$\beta\varepsilon\rightarrow \infty$ we have 
$\langle X^2\rangle\simeq p(1-p)R^2$ like for non-fluctuating 
interspacing while 
$\langle X^2\rangle\simeq 3p(1-p){\rm k_BT}/2\sigma$ as 
$\beta\varepsilon\rightarrow 0$ like for motions in a harmonic potential. 
This behavior is compatible and originates from jump dynamics between 
soft walls, the broadening of the point sites with temperature resulting 
from the environment fluctuations.

Generalization of Eqs.(\ref{ACjump}) and (\ref{x2jump}) to the case of 
jump dynamics among $N$ equivalent sites on a fluctuating circle of radius 
$a$ is straightforward. In this case, the normalized scattering intensity still 
writes  like in  Eq.(\ref{IQ0}) with the EISF given by 
$A(Q)=\langle A_0(Qa)\rangle$ and the $Q$-dependent scattering correlation 
function by 
\begin{eqnarray}
C(Q,t) & = & \left[\sum_{l=1}^{N}\frac{\langle A_l(Qa)\rangle}
{1-\langle A_0(Qa)\rangle}\right.\, \nonumber \\
& \times & \left.
\exp\left\{-\left(\frac{2t}{\tau}\right)\sin^2\left(\frac{l\pi}{N}\right)\right\}
\right]\,\psi(t)\:,
\end{eqnarray}
where the amplitude $A_l(Qa)=N^{-1}\sum_{n=1}^{N}j_0\left[2Qa\sin(n\pi/N)\right]
\cos(2ln\pi/N)$ for $l=0, 1, \cdots, N$, and $\tau$ is the effective mean 
residence time in each site. Interestingly, the mean-squared displacement 
of the particle is simply $\langle X^2\rangle=\langle a^2\rangle$ where 
$\langle a^2\rangle$ is obtained from Eq.(\ref{x2jump}).

\section*{Fluctuating Sphere}

\begin{figure}[ht]
\vspace{0.3cm}
\centerline{\hbox{
\hspace{-0.3cm}
\psfig{figure=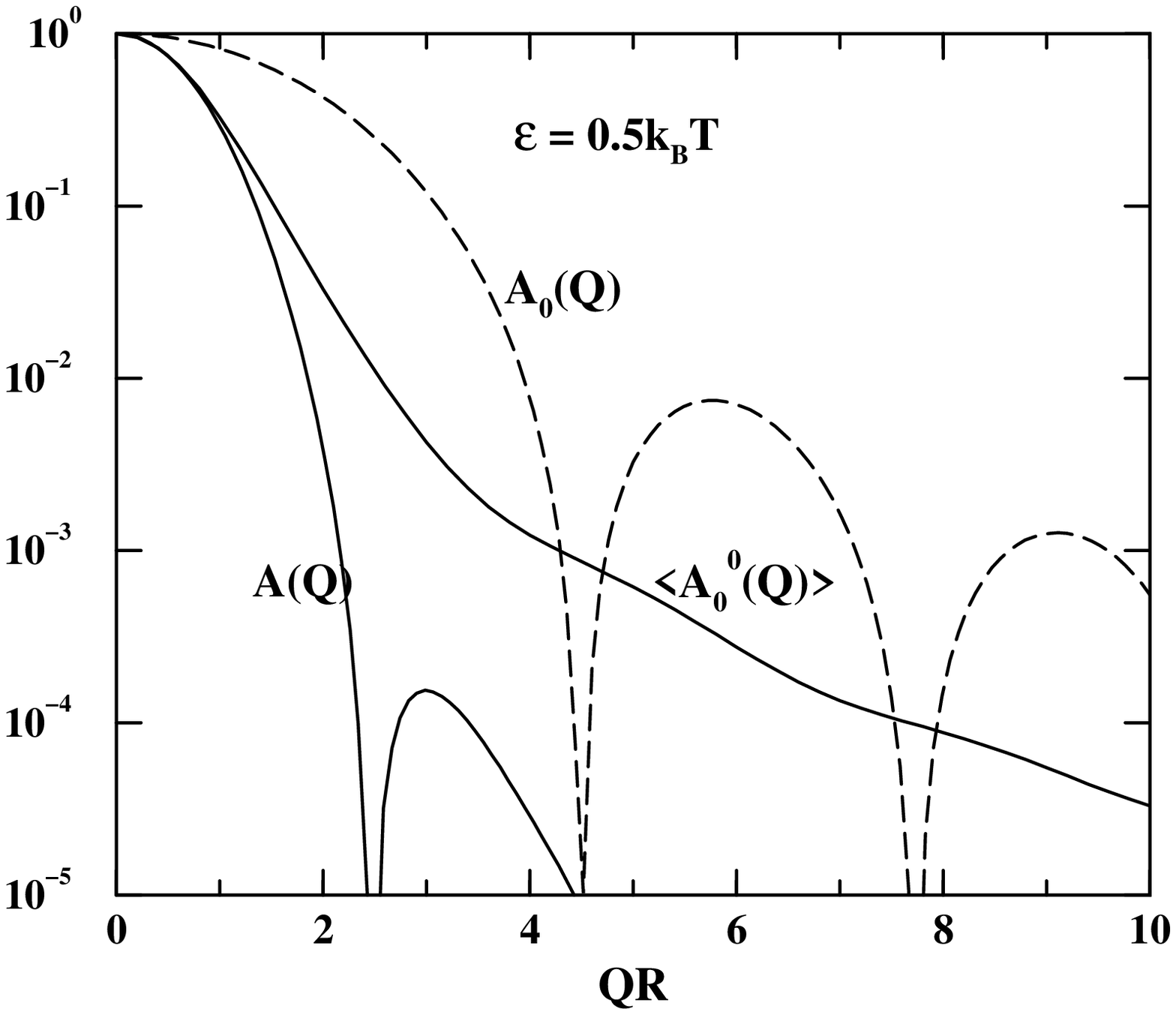,width=3.2in,angle=0}
\hspace{0.3cm}
\psfig{figure=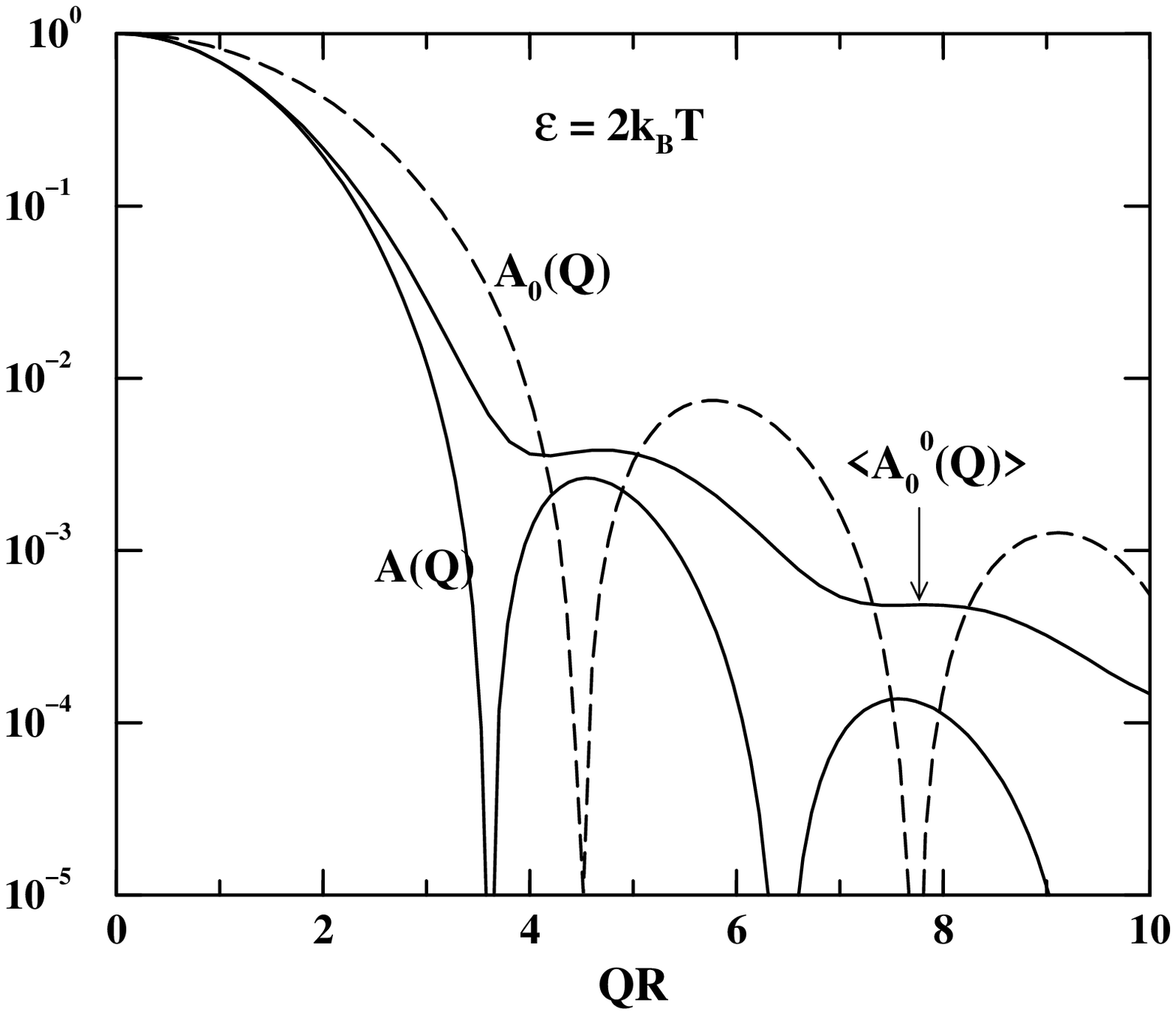,width=3.2in,angle=0}
}}
\vspace{0.25cm}
\caption{The amplitudes $\langle A_0^0(Q)\rangle$ and $A(Q)$ in 
Eqs.(\ref{A0Q}) and (\ref{AQ}), respectively, as a function of $QR$ for 
two values of the reduced energy $\varepsilon/{\rm k_BT}=0.5,2$ as quoted 
on the figures. The dashed lines represent $A_0(Q)$ in Eq.(\ref{FQ}) for 
$a=R$ for comparison. Note that $QR\in [0.1,6]$ in the typical neutron 
scattering experiment.}
\label{eisfs}
\end{figure}

\begin{figure}[ht]
\vspace{0.3cm}
\centerline{\hbox{
\hspace{-0.3cm}
\psfig{figure=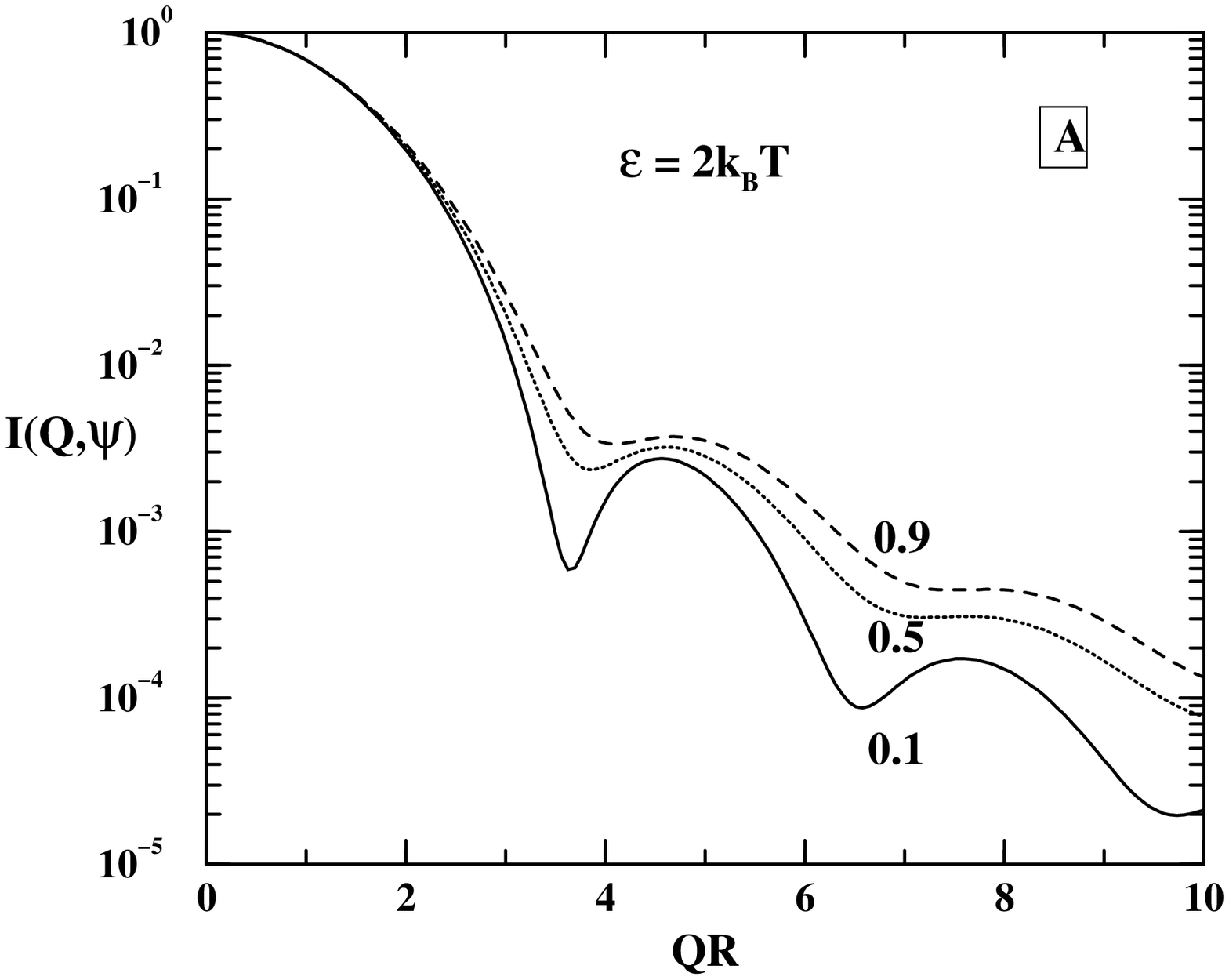,width=3.22in,angle=0}
\hspace{0.3cm}
\psfig{figure=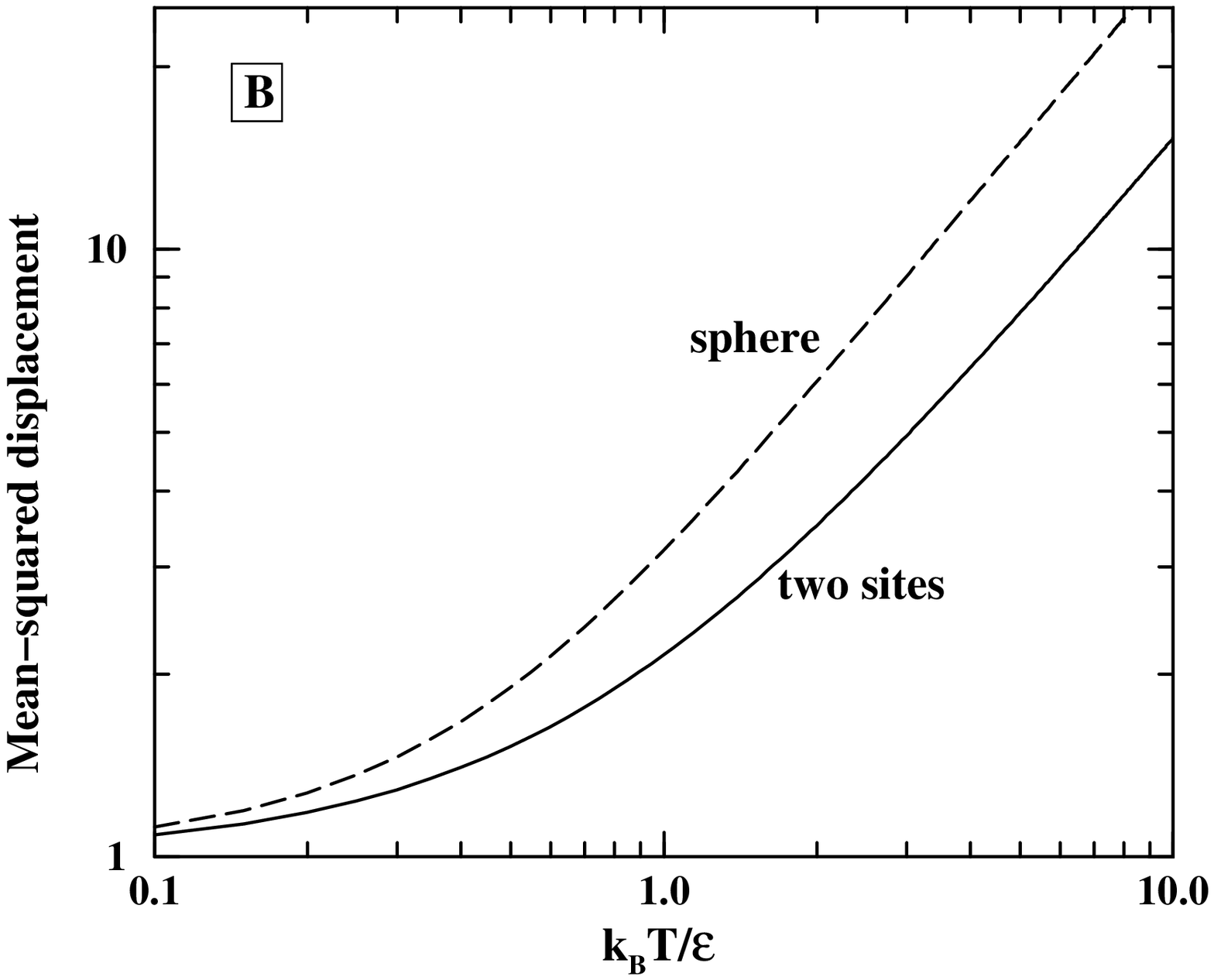,width=3.2in,angle=0}
}}
\vspace{0.25cm}
\caption{{\bf Panel A:} The intensity $I(Q,\psi)$ in Eq.(\ref{IQ1}) 
for diffusion inside a fluctuating sphere, versus $QR$ for 
$\psi(\Gamma)=0.1,0.5,0.9$ as quoted on the figure. 
{\bf Panel B:} Reduced mean-squared displacements 
$\langle X^2\rangle/p(1-p)R^2$ in Eq.(\ref{x2jump}) (solid line) for jump 
dynamics and $5\langle X^2\rangle/3R^2$ in Eq.(\ref{x2}) 
(dashed line) for diffusion, as a function of the reduced temperature 
${\rm k_BT}/\varepsilon$.}
\label{roft}
\end{figure}

The potential for diffusion inside a sphere of radius $a$ with impermeable 
surface is 
\begin{eqnarray}
U(r,a)=\left\{\begin{array}{ccl}
0 & ; & r\leq a\:,\\
\infty & ; & r>a\:.
\end{array}\right.
\label{pots}
\end{eqnarray}
In this case, the normalized scattering intensity is given by Eq.(\ref{IQ0}) 
with the scattering correlation function given by 
\begin{eqnarray}
C(Q,t) & = & \left\{\sum_{n,l=0}^{\infty}\int_{0}^{\infty}\!da\,\right.
\underbrace{p_{\rm eq}(a)\,\left[\frac{\left[A_n^l(Qa)-A(Q)\delta_{0l}\,
\delta_{0n}\right]}{\left[1-A(Q)\right]}\right]}_{{\overline g}_n^l(Qa,a)}\:
 \nonumber \\
& & \left. \times\:{\rm e}^{-(x_n^l)^2Dt/a^2}\right\}\,\psi(t)\:,
\label{CQts}
\end{eqnarray}
where the $A_n^l(Qa)$ are given by \cite{Vol,Bic}
\begin{equation}
A_n^l(Qa)=\frac{6(2l+1)(x_n^l)^2}{(x_n^l)^2-l(l+1)}\,
\left[\frac{Qa\,j_{l+1}(Qa)-lj_{l}(Qa)}{(Qa)^2-(x_n^l)^2}\right]^2\:.
\end{equation}
${\overline g}_n^l(Qa,a)$ is the $Q$-dependent effective distribution of 
relaxation times $\tau_n^l(a)=a^2/(x_n^l)^2D$, $D$ is the effective diffusion 
constant and $x_n^l$ are the roots of equation \cite{Vol,Bic}, 
$x_n^l\,j_{l+1}(x_n^l)=lj_{l}(x_n^l)$.  For the potential in Eq.(\ref{pot}), 
the structure factors $\langle A_0^0(Q)\rangle$ (i.e., the 
EISF for a sphere of radius $a$ averaged over the distribution of radii) 
and the EISF of the system $A(Q)$ are therefore: 
\setcounter{xeq}{\value{equation}} 
\addtocounter{xeq}{1}
\renewcommand{\theequation}{\arabic{xeq}\alph{equation}}
\setcounter{equation}{0}
\begin{eqnarray}
\langle A_0^0(Q)\rangle=\langle|F(Q)|^2\rangle & = & 
\frac{9 \int_{1}^{\infty}x^3\,[j_1(xQR)]^2\,{\rm e}^{-\beta\varepsilon x^2}\,dx}
{(QR)^2 \int_{1}^{\infty}x^5\,{\rm e}^{-\beta\varepsilon x^2}\,dx}\:, 
\label{A0Q}  \\
A(Q)=|\langle F(Q)\rangle|^2 & = & \left|
\frac{3\int_{1}^{\infty}x^4\,j_1(xQR)\,{\rm e}^{-\beta\varepsilon x^2}\,dx}
{QR \int_{1}^{\infty}x^5\,{\rm e}^{-\beta\varepsilon x^2}\,dx}\right|^2\:.  
\label{AQ}
\end{eqnarray}
\setcounter{equation}{\value{xeq}} 
\renewcommand{\theequation}{\arabic{equation}}
Figure \ref{eisfs} shows that $A(Q)\leq \langle A_0^0(Q)\rangle$ for 
all $QR$, and the difference between the two functions increases 
with increasing the temperature, i.e. as $\beta\varepsilon$ gets smaller. 
The location of minima of $A(Q)$ and $\langle A_0^0(Q)\rangle$ do not 
coincide and their values are equal to zero for $A(Q)$ while they are 
different from zero for $\langle A_0^0(Q)\rangle$. Like in Fig.~\ref{Ajump}, 
the location of minima are shifted to the left compare to $A_0(Q)$ 
and tend to disappear as $\beta\varepsilon$ gets smaller. To assess 
to which extent $A(Q)$ and $\langle A_0^0(Q)\rangle$ contribute to the 
elastic intensity, we turn back to Eq.(\ref{CQts}). The 
shortest relaxation time of this expansion is $\Gamma^{-1}$, where 
$\Gamma=(x_1^0)^2D/R^2$ is the typical time for a particle to diffuse 
over the entire sphere of radius $R$. When $t\sim \Gamma^{-1}$, the second 
term containing the exponential time-dependence in Eq.(\ref{CQts}) 
can be neglected. In this case the incoherent intermediate scattering 
function reduces to:
\begin{equation}
I(Q,\psi)=\psi(\Gamma)\,\langle A_0^0(Q)\rangle
+\left[1-\psi(\Gamma)\right]\,A(Q)\:.
\label{IQ1}
\end{equation}
This expression, which would represent the elastic part of $S(Q,\omega)$ for 
the resolution time of about $\Gamma^{-1}$, contrasts with the case in the 
absence of fluctuations where the incoherent intermediate scattering 
function relaxes to the EISF for time scales of order or greater than 
$\Gamma^{-1}$. Figure~\ref{roft}A displays $I(Q,\psi)$ versus $QR$ for 
various values of $\psi(\Gamma)$. Manifestly, $I(Q,\psi)$ is different 
from the EISF $A(Q)$ even for $\psi(\Gamma)=0.1$. It is obvious that 
$I(Q,\psi)$ will eventually be equal to the EISF in the 
$\psi(\Gamma)\rightarrow 0$ limit when the particle diffusion is very 
slow compare to fluctuations of the sphere radius. In the opposite limit 
when the particle diffusion is fast compare to sphere radius fluctuations, i.e. 
$\psi(\Gamma)\sim 1$, the $I(Q,\psi)$ is essentially given by 
$\langle A_0^0(Q)\rangle$. The mean-squared displacement of the particle is
\begin{eqnarray}
& & \langle X^2\rangle=-3\left.\frac{d\ln\left[I(Q,\psi)\right]}{d(Q^2)}
\right|_{Q=0}=\langle r^2\rangle \nonumber \\
& & =\frac{3R^2}{5}
\left\{\frac{3}{\beta\varepsilon}+
\left[1+\frac{2}{\beta\varepsilon}+
\frac{2}{(\beta\varepsilon)^2}\right]^{-1}\right\}\:. 
\label{x2}
\end{eqnarray}
Note that $\langle X^2\rangle$ is independent of $\psi(\Gamma)$ since, as 
is illustrated in Fig.~\ref{eisfs}, in the Gaussian scattering approximation 
the structure factors are, $A(Q)=A_0^0(Q)\simeq 1-3Q^2\langle r^2\rangle/5$ 
as $Q\rightarrow 0$. Figure~\ref{roft}B shows the temperature 
dependence of the particle mean-squared displacement. At low temperature 
$\langle X^2\rangle=3R^2/5$ like for diffusion inside a sphere of radius 
$R$ while $\langle X^2\rangle\sim 9{\rm k_BT}/5\sigma$ at higher temperature 
like for motions in a harmonic potential. This behavior is compatible and 
originates from diffusion inside a sphere with a thick soft surface, the 
thickening of the spherical surface with temperature resulting from 
the environment fluctuations. 
 
Let us turn now to the quasielastic term of $S(Q,\omega)$. The dynamics 
involved can be characterized by considering the position correlation 
function $C(0,t)$ obtained in taking the $Q\rightarrow 0$ limit of  
$C(Q,t)$ in Eq.(\ref{CQts}) to give: 
\begin{eqnarray}
& & C(0,t)=\left[\int_{1}^{\infty}\!{\overline g}(\beta\varepsilon,x)\,
{\rm e}^{-\Gamma t/x^2}\,dx\right]\,\psi(t) \label{corr} \\
& & {\overline g}(\beta\varepsilon,x)=
\frac{x^7\,{\rm e}^{-\beta\varepsilon x^2}}
{\int_{1}^{\infty}\!x^7\,{\rm e}^{-\beta\varepsilon x^2}\,dx}\:.
\nonumber 
\end{eqnarray}
The effective distribution ${\overline g}(\beta\varepsilon,x)$ of relaxation 
times $x^2/\Gamma$ is maximum at $x_m=\rm {max}[1,(7/2\beta\varepsilon)^{1/2}]$. 
For the purpose of illustration we consider the situation where the 
fluctuation dynamics of the sphere radius is described by the Langevin 
oscillator with the correlation function
\begin{eqnarray}
& & \psi(t)={\rm e}^{-\gamma t/2}\,\left[{\rm cosh}(\mu t)+
\left(\frac{\gamma}{2\mu}\right)\,{\rm sinh}(\mu t)\right]
\label{radcor1} \nonumber \\
& & \mu=\frac{\sqrt{\gamma^2-4\omega_0^2}}{2}\:,
\end{eqnarray}
where $\gamma$ and $\omega_0$ are the collision frequency (related to 
the dissipation) and the frequency of the oscillator. The initial decay 
rate constant of $\psi(t)$ is zero and its relaxation time is 
$\gamma/\omega_0^2$. 

\begin{figure}[ht]
\vspace{0.3cm}
\centerline{\hbox{
\hspace{-0.3cm}
\psfig{figure=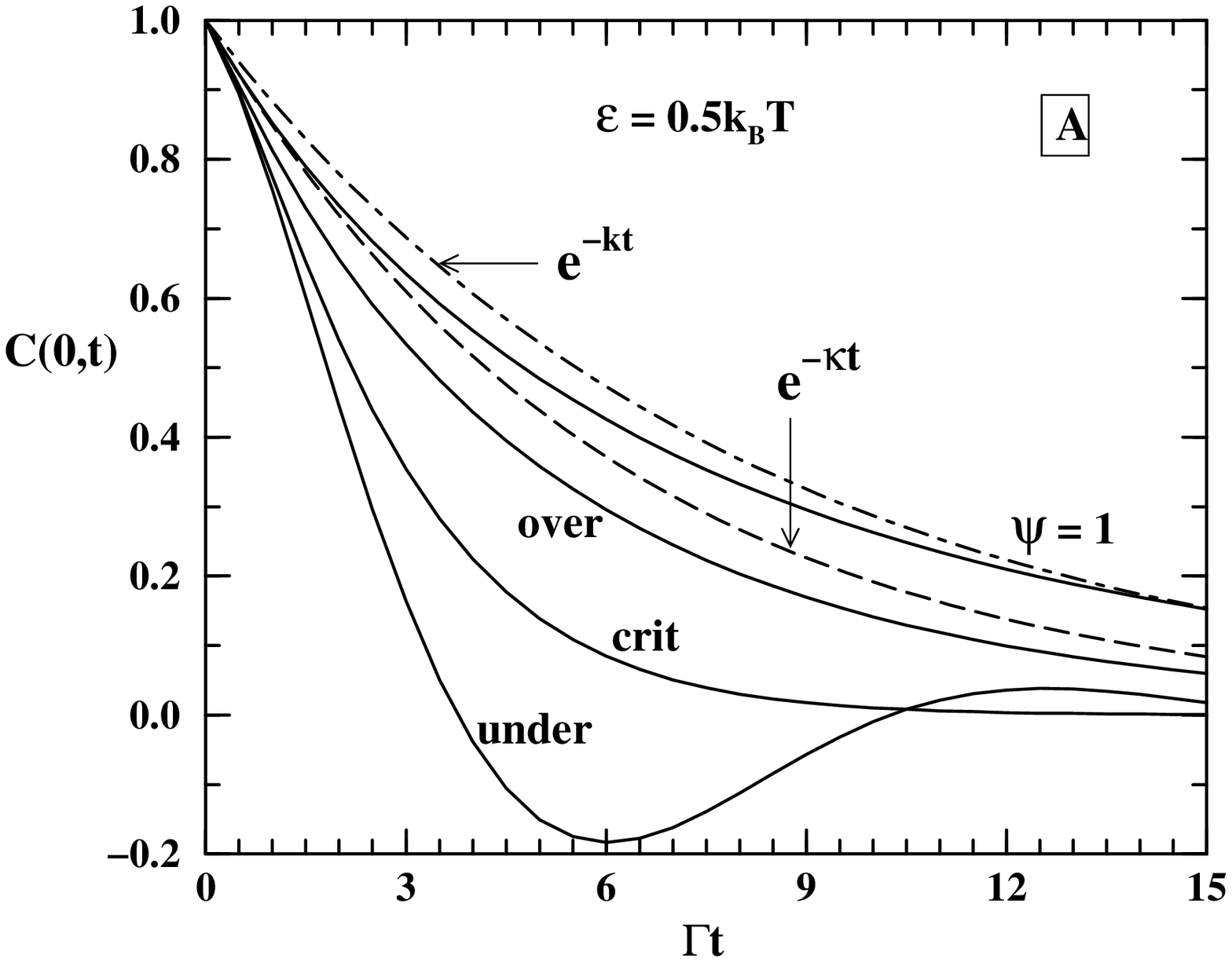,width=3.2in,angle=0}
\hspace{0.3cm}
\psfig{figure=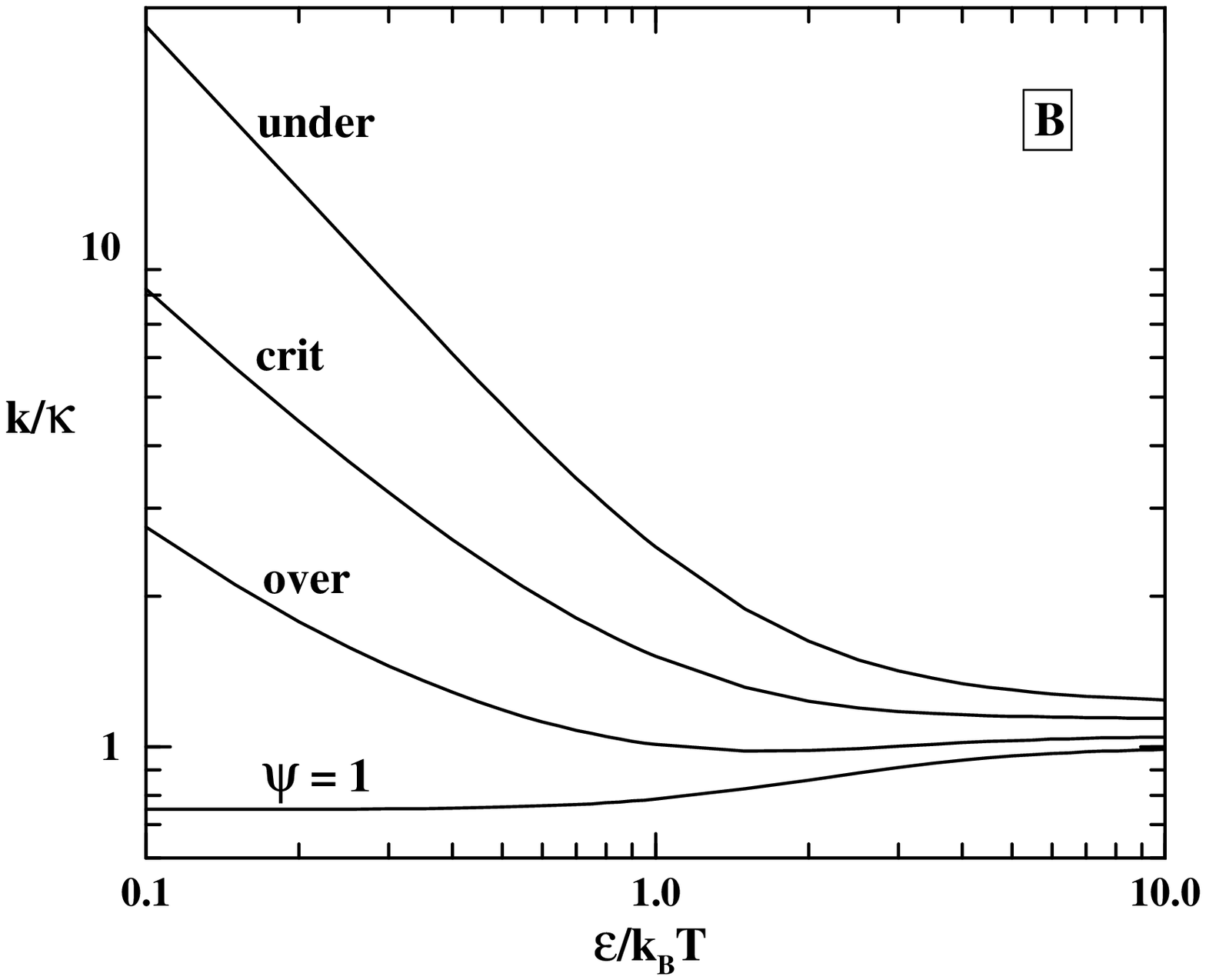,width=3.2in,angle=0}
}}
\vspace{0.25cm}
\caption{{\bf Panel A:} Correlation function $C(0,t)$ in Eq.(\ref{corr}) 
as a function of the reduced time $\Gamma t$ for the reduced energy 
$\beta\varepsilon=0.5$. The dashed and dot-dashed lines represent 
${\rm e}^{-\kappa t}$ and ${\rm e}^{-kt}$, respectively, with 
$\kappa=13\Gamma/79$ and $k=79\Gamma/633$. 
{\bf Panel B:} Reduced relaxation rate $k/\kappa$ (ratio of Eqs.(\ref{rate}) 
and (\ref{kap})) as a function of $\varepsilon/{\rm k_BT}$. For the two panels,
we have $\omega_0=\Gamma/2$ and the quotations 'under' ($\gamma=\Gamma/4$), 
'crit'($\gamma=\Gamma$), and 'over' ($\gamma=4\Gamma$) correspond 
respectively to underdamped, critical and overdamped regimes of $\psi(t)$ in 
Eq.(\ref{radcor1}).} 
\label{coft}
\end{figure}

As shown in Fig.~\ref{coft}A, because of the distribution of relaxation 
times due to fluctuations of the sphere radius the $C(0,t)$ is no 
longer a single exponential. When $\psi(t)=1$, the short time behavior of 
$C(0,t)$ can be fitted by ${\rm e}^{-\kappa t}$, where $\kappa$ is the 
initial decay rate (see below) of $C(0,t)$, while ${\rm e}^{-kt}$, where 
$k$ is the relaxation rate (see below) of $C(0,t)$, is a poor approximation 
of $C(0,t)$ although they coincide around $\Gamma t=15$.  
In the presence of fluctuations when $\psi(t)\neq 1$, $C(0,t)$ is 
essentially dominated by $\psi(t)$. 

The initial decay rate of $C(0,t)$ is,
\begin{eqnarray}
\kappa & = & -\left.\frac{dC(0,t)}{dt}\right|_{t=0}=
\frac{3R^2}{5\langle r^{2}\rangle}\,\Gamma\:, \label{kap}\\ 
& \simeq & \left\{\begin{array}{lcc}
\Gamma & ; & \beta\varepsilon\rightarrow \infty\:\:\mbox{(low T)}\:,\\
\beta\varepsilon\Gamma/3 & ; & \beta\varepsilon\rightarrow 0
\:\:\mbox{(high T)}\:.
\end{array}\right. \nonumber 
\end{eqnarray}
In this example, $\kappa$ is independent of $\psi(t)$ and $\Gamma/\kappa$ 
coincides with the mean-squared displacement. If one uses for 
the diffusion coefficient the relation $D={\rm k_BT}/m\xi$, 
where $m$ is the particle mass and $\xi$ the friction coefficient, we 
find that $\kappa\propto {\rm T}$ at low ${\rm T}$ while 
$\kappa\simeq (x_1^0)^2\sigma/m\xi$ independent of temperature for high 
${\rm T}$. 

The relaxation rate, $k^{-1}=\int_{0}^{\infty}\!C(0,t)\,dt$, which 
depends on characteristics of $\psi(t)$, is 
\begin{eqnarray}
\frac{\Gamma}{k} & = & \left[\int_{1}^{\infty}\!x^7\,
{\rm e}^{-\beta\varepsilon x^2}\,dx\right]^{-1} \nonumber \\
& \times & \int_{1}^{\infty}\!dx\,\left[
\frac{x^9\,(1+\gamma x^2/\Gamma)\,{\rm e}^{-\beta\varepsilon x^2}}
{1+\gamma x^2/\Gamma+(\omega_0/\Gamma)^2x^4}\right]\:.
\label{rate}
\end{eqnarray}
The ratio $k/\kappa$ as a function of inverse temperature is plotted in 
Fig.~\ref{coft}B. Three types of decays of $C(0,t)$ can be noted: $k/\kappa<1$ 
the $C(0,t)$ has a 
nonexponential decay dominated by the distribution of relaxation times, 
$k/\kappa=1$ then $C(0,t)={\rm e}^{-\kappa t}$, and $k/\kappa>1$ where 
the nonexponential $C(0,t)$ (which originates from both the distribution 
of relaxation times and $\psi(t)$) may show oscillations and have long tail 
decay. These differences in the decay of $C(0,t)$ are more pronounced 
for $\beta\varepsilon\rightarrow 0$ (high T) where $k/\kappa$ can be 
very large while $k/\kappa\rightarrow 1$ (i.e., $C(0,t)$ is almost a 
single exponential) when $\beta\varepsilon\rightarrow \infty$ (low T). 
Finally, it goes without saying that when we are concern with the jump 
diffusion in a fluctuating sphere, all the expressions derived above remain 
unchanged except that the $\exp\{-(x_n^l)^2Dt/a^2\}$ in Eq.(\ref{CQts}) 
is replaced by 
$\exp\left\{-\left[1-{\rm e}^{-(x_n^l b)^2/2a^2}\right]\,\gamma_j t\right\}$, 
where $b$ and $\gamma_j$ are the jump length and frequency \cite{Bic}, 
respectively. 

\section*{Summary}

Let us summarize the main results derived above. We find that as a result 
of the environment fluctuations the EISF of the system is given by 
Eq.(\ref{AofQ}) and the incoherent scattering correlation function $C(t)$, 
given in Eq.(\ref{CQt}), is 
$C(Q,t)=\sum_{n}\langle{\rm e}^{-\lambda_n(a) t}\rangle\,\psi(t)$, 
where the average is taken over the $Q$-dependent effective distribution 
of relaxation times and $\psi(t)$ is the correlation function of the 
fluctuating length. 

For the jump dynamics between sites, Eq.(\ref{AofQ}) reduces 
to $\langle A_0(Qa)\rangle$ which is the EISF $A_0(Qa)$ (in Eq.(\ref{A0j})) 
in the absence of fluctuations averaged over the distribution of site 
interspacing. As the relaxation rate $\Gamma$ (in the absence of 
fluctuations) is independent of the site interspacing, the incoherent scattering 
correlation function $C(t)$, given in Eq.(\ref{ACjump}), is equal to 
$C(t)$ in the absence of fluctuations times $\psi(t)$. 

For the diffusion inside a fluctuating sphere, on the other hand, Eq.(\ref{AofQ}) 
reduces to $\left|\langle F(Qa)\rangle\right|^2$ where $F(Qa)$, given 
in Eq.(\ref{FQ}), is the scattered amplitude for diffusion inside a sphere 
of fixed radius $a$ and the average is taken over the distribution of radii. 
However, the normalized elastic intensity $I(Q,\psi)$, given by Eq.(\ref{IQ1}), 
involves two contributions depending on the relaxation of the sphere radius 
over time scales $\Gamma^{-1}$. When the particle diffusion is slow compare 
to the fluctuations of the sphere, $I(Q,\psi)$ is equal to 
$\left|\langle F(Qa)\rangle\right|^2$ while, in the opposite limit, it is 
is given by $\langle\left|F(Qa)\right|^2\rangle$. Since the relaxation time 
$\tau_n^l(a)$ for the diffusion in a sphere depend on the sphere radius, 
the fluctuations of the sphere radius generate an effective distribution 
of relaxation time so that $C(Q,t)$, in Eq.(\ref{CQts}), is 
$C(Q,t)=\sum_{n,l=0}^{\infty}\langle{\rm e}^{-t/\tau_n^l(a)}\rangle\,\psi(t)$, 
where the average is taken over the $Q$-dependent effective distribution 
of relaxation times and $\psi(t)$ is the correlation function of the sphere 
radius. 

It worthwhile to mention that when $\psi(t)=1$, the situation resembles to 
what is known as the heterogeneous scenario for the explanation of the 
stretched exponential or Kohlrausch-Williams-Watts relaxation \cite{Arbe}. 
Such a situation may be encountered in the case where the particle atoms in 
a protein system undergo jump dynamics between sites (or diffusion inside 
spherical cages) each of them with different site interspacing (radius), i.e. 
polydispersity in length scales. In this case, $C(Q,t)$ is exponential for 
jump dynamics between sites (provided that $\Gamma$ is the same for each 
particle) while it is nonexponential for diffusion inside a sphere. The above 
results with $\psi(t)\neq 1$ are derived in the homogeneous 
scenario \cite{Arbe} which assumes that all particles are dynamically identical. 

To conclude, we emphasize that the dynamics of particle atoms in a protein 
system is a multidimensional problem in which the particles and constituent 
elements of their environment undergo their own dynamics in respective potentials. 
The speculative discussion outlined above is a quite simplified two-dimensional 
version (particle position ${\bf r}$ and site interspacing or sphere radius $a$) 
of the problem. Nonetheless, it is encouraging to see that the present 
analysis indicate that the form of the scattering intensity, for instance, is 
quite influenced by the fluctuations. This suggests at least to reconsider the 
way of fitting experimental data and revise the interpretation of parameters as 
determined. In the same spirit, an analysis combining particle dynamics and 
fluctuations in both the size and shape of the confining geometries can be 
developed along these lines.

\subsection*{Acknowledgments} 
Thanks are due to J. Dianoux and E. Kats for their comments on the 
manuscript.


\onecolumn

\vspace{0.25cm}

\vspace{0.25cm}

\begin{thebibliography}{99} 

\bibitem{Bee} M. B\'ee, {\it Quasielastic Neutron Scattering}, (Adam Hilger, 
Bristol, 1988).

\bibitem{Vol} F. Volino and A. J. Dianoux, Mol. Phys. \textbf{41}, 271 (1980). 

\bibitem{Hall}  P. L. Hall and D. K. Ross, Mol. Phys. \textbf{42}, 673 (1981). 

\bibitem{Bel} M.-C. Bellissent-Funel, J. Teixeira, K. F. Bradley 
and S. H. Chen, J. Phys. I (France) \textbf{2}, 995 (1992); 
M.-C. Bellissent-Funel, J-M. Zanotti and S. H. Chen, 
Faraday Discuss. \textbf{103}, 281 (1996).  

\bibitem{BelB} {\it Hydration Processes in Biology: Theoretical and 
Experimental Approaches}, Ed. M.-C. Bellissent-Funel, 
(IOS press, Amsterdam, 1999, Vol. 305). 

\bibitem{Doster} W. Doster, S. Cusack and W. Petry, 
Nature \textbf{337}, 754 (1989).

\bibitem{Fitter} J. Fitter, R. E. Lechner, G. B\"{u}ldt, and 
N. A. Dencher, Proc. Natl. Acad. Sci. USA \textbf{93}, 7600 (1996). 

\bibitem{Per} J. Perez, J.-M. Zanotti, and D. Durand, 
Biophys. J. \textbf{77}, 454 (1999).

\bibitem{Den} In {\it Hydration Processes in Biology: Theoretical and 
Experimental Approaches}, Ed. M.-C. Bellissent-Funel, 
(IOS press, Amsterdam, 1999, Vol. 305) pp. 195 - 217.

\bibitem{Gotze} W. G\"{o}tze and L. Sj\"{o}gren. 
Rep. Prog. Phys. \textbf{55}, 241 (1992).

\bibitem{Lee} B. Lee, Proc. Natl. Acad. Sci. USA (Chemistry) 
\textbf{80}, 622 (1983).

\bibitem{Bic} D. J. Bicout, Phys. Rev. E. \textbf{62}, 261 (2000). 

\bibitem{Arbe} A. Arbe, J. Colmenero, M. Monkenbusch and D. Richter, 
Phys. Rev. Lett. \textbf{81}, 590 (1998). 


\end{thebibliography}
\end{document}